\documentclass[11pt,twoside]{article}
\usepackage{asp2010}
\newcommand{\maa}{\alpha \alpha} 
\newcommand{\mga}{\gamma \alpha} 

\resetcounters

\begin{document}

\title{Progenitors of Supernovae Type Ia}
\author{Silvia Toonen,$^{1,\star}$ Gijs Nelemans,$^{1,2}$ Madelon Bours,$^{1,3}$ Simon Portegies Zwart,$^4$ Joke Claeys,$^{1,5}$ Nicki Mennekens,$^6$ and Ashley Ruiter$^7$
\affil{$^1$ Department of Astrophysics/IMAPP, Radboud University Nijmegen, P.O. Box 9010, NL-6500 GL, Nijmegen, the Netherlands; $\star$ email: {\tt silviato@astro.ru.nl}}
\affil{$^2$ Instituut voor Sterrenkunde, KU Leuven, Celestijnenstraat 200D, 3001 Leuven, Belgium}
\affil{$^3$ Department of Physics, University of Warwick, Coventry CV4 7AL, United Kingdom }
\affil{$^4$ Leiden Observatory, Leiden University, P.O. Box 9513, NL-2300 RA,  Leiden, The Netherlands}
\affil{$^5$ Sterrekundig Instituut Utrecht, PO Box 80000, 3508 TA Utrecht, The Netherlands}
\affil{$^6$ Astrophysical Institute, Vrije Universiteit Brussel, Pleinlaan 2, 1050 Brussels, Belgium}
\affil{$^7$ Max Planck Institute for Astrophysics, Karl-Schwarzschild-Str. 1, 85741 Garching, Germany}
}

\begin{abstract}
Despite the significance of Type Ia supernovae (SNeIa) in many fields in astrophysics, SNeIa lack a theoretical explanation. The standard scenarios involve thermonuclear explosions of carbon/oxygen white dwarfs approaching the Chandrasekhar mass; either by accretion from a companion or by a merger of two white dwarfs. We investigate the contribution from both channels to the SNIa rate with the binary population synthesis (BPS) code SeBa in order to constrain binary processes such as the mass retention efficiency of WD accretion and common envelope evolution. We determine the theoretical rates and delay time distribution of SNIa progenitors and in particular study how assumptions affect the predicted rates. 
\end{abstract}

\section{Method}
Binary population synthesis (BPS) codes are very useful tools to study the evolution of binary stars and the processes that govern them. They enable the simulation of large numbers of binary systems and with that the study of populations of e.g. novae, X-ray binaries and Type Ia supernova (SNeIa). BPS codes are particularly useful to study trends in stellar populations and the effects that various important (and uncertain) phases have on the binary evolution.
Here we investigate the population of SNIa progenitors from two possible evolutionary channels using the BPS code SeBa\footnote{SeBa is incorporated in the Astrophysics Multipurpose Software Environment, or AMUSE. This is a component library with a homogeneous interface structure, and can be downloaded for free at {\tt amusecode.org} \citep{Por09}.} \citep{Por96, Nel01, Too12}. In the single degenerate (SD) channel \citep{Whe73} a carbon/oxygen white dwarf (WD) accretes from a hydrogen-rich companion to approach the Chandrasekhar mass. The double degenerate (DD) channel describes the merger of two carbon/oxygen white dwarfs \citep{Ibe84, Web84}. Other channels are a.o. accreting WDs from helium-rich non-degenerate companions \citep[][see e.g.]{Wan09b} and the merger between a WD and the core of an asymptotic giant branch star \citep{Kas11}. We study the SNIa delay time distribution (DTD), where the delay time is the time between the formation of the binary system and the SNIa event. In a simulation of a single burst of star formation the DTD gives the SNIa rate as a function of time after the starburst. The DTD is linked to the nuclear timescales of the progenitors and the binary evolution timescales up to the merger.

\section{Double degenerate channel}
We set out to predict SNIa rates for the DD channel, with the additional constraint that our model corresponds well to the observed population of close double WDs (CDWDs) - of all masses and compositions. Even though there are no certain DD SNIa progenitors among the observed CDWDs e.g \citet{Mar11b}, the DD SNeIa and observed CDWD progenitors have gone through similar evolutionary paths and are affected by the same binary and stellar processes. 

\begin{figure*}
    \centering
    \begin{tabular}{c c}
	\includegraphics[scale=0.3, angle=270]{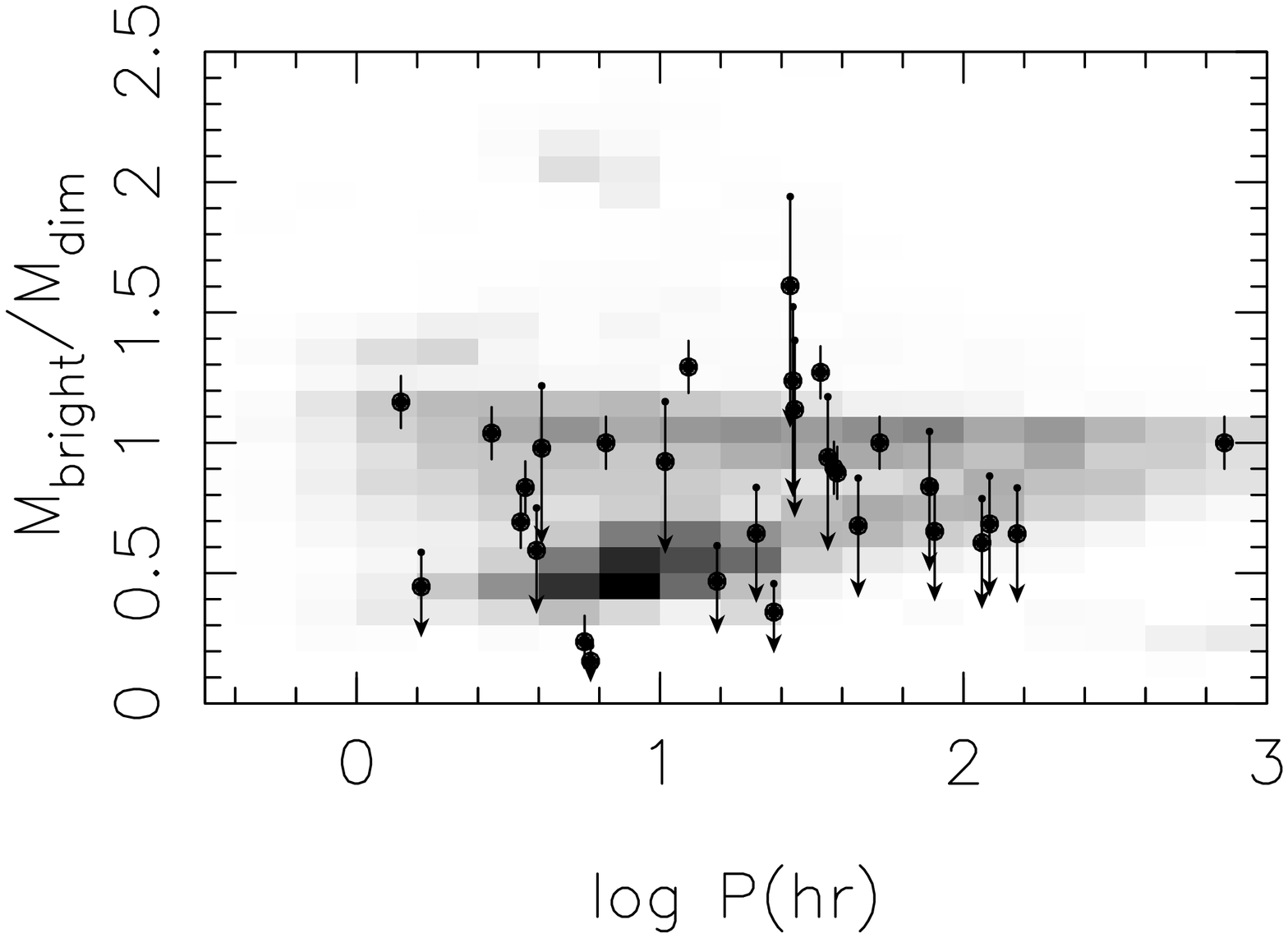} &
	\includegraphics[scale=0.3, angle=270]{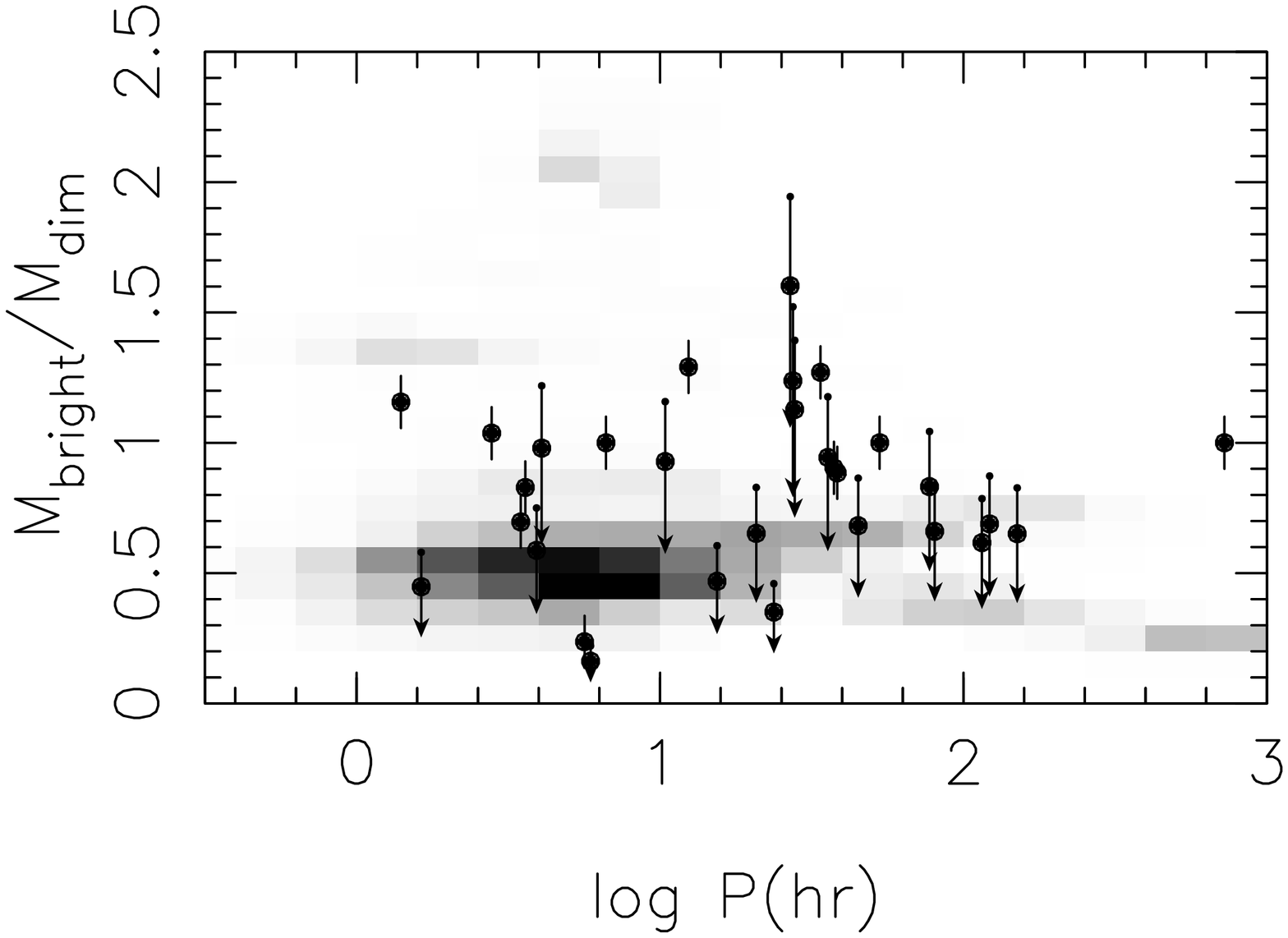} \\
	(a) & (b) \\
	\end{tabular}
    \caption{\small Simulated distribution of population of visible double white dwarfs as a function of orbital period and mass ratio, where mass ratio is defined as the mass of the brighter white dwarf divided by that of the dimmer white dwarf. Left model $\mga$ is used, on the right model $\maa$. The intensity of the grey scale corresponds to the density of objects on a linear scale. The same grey scale is used for both plots. Observed binary white dwarfs \citep{Mar11b} are overplotted with filled circles.} 
    \label{fig:pop_dwd}
    \end{figure*}

The CDWDs are believed to engage in at least one phase of common envelope (CE) evolution. In spite of the importance of the CE phase, it remains poorly understood. Several prescriptions for CE evolution have been proposed. The $\alpha$ formalism \citep{Web84} is based on the conservation of orbital energy and the $\gamma$-formalism \citep{Nel00} is based on the conservation of angular momentum. In model $\alpha\alpha$ the $\alpha$-formalism is used to determine the outcome of every CE. For model $\gamma\alpha$ the $\gamma$-prescription is applied unless the binary contains a compact object or the CE is triggered by the Darwin-Riemann instability \citep{Dar1879}. The $\gamma\alpha$-model reproduces the mass ratio distribution of the observed DWDs best, see Fig. \ref{fig:pop_dwd}. 

\begin{figure}[t]
\begin{tabular}{c c}
\includegraphics[width=0.5\columnwidth, trim = 0 20 0 500]{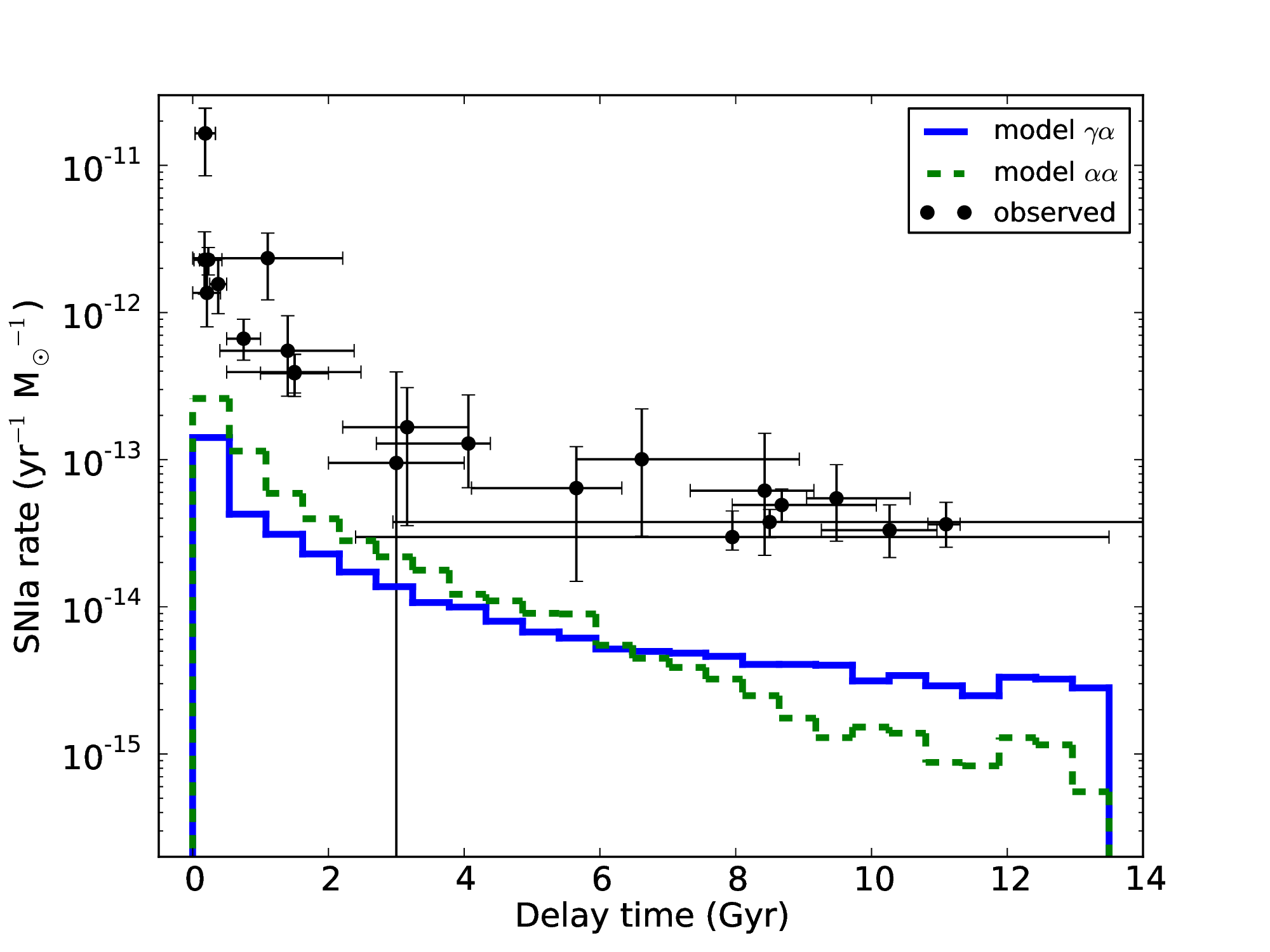} & 
\includegraphics[width=0.5\columnwidth, trim = 0 20 0 500]{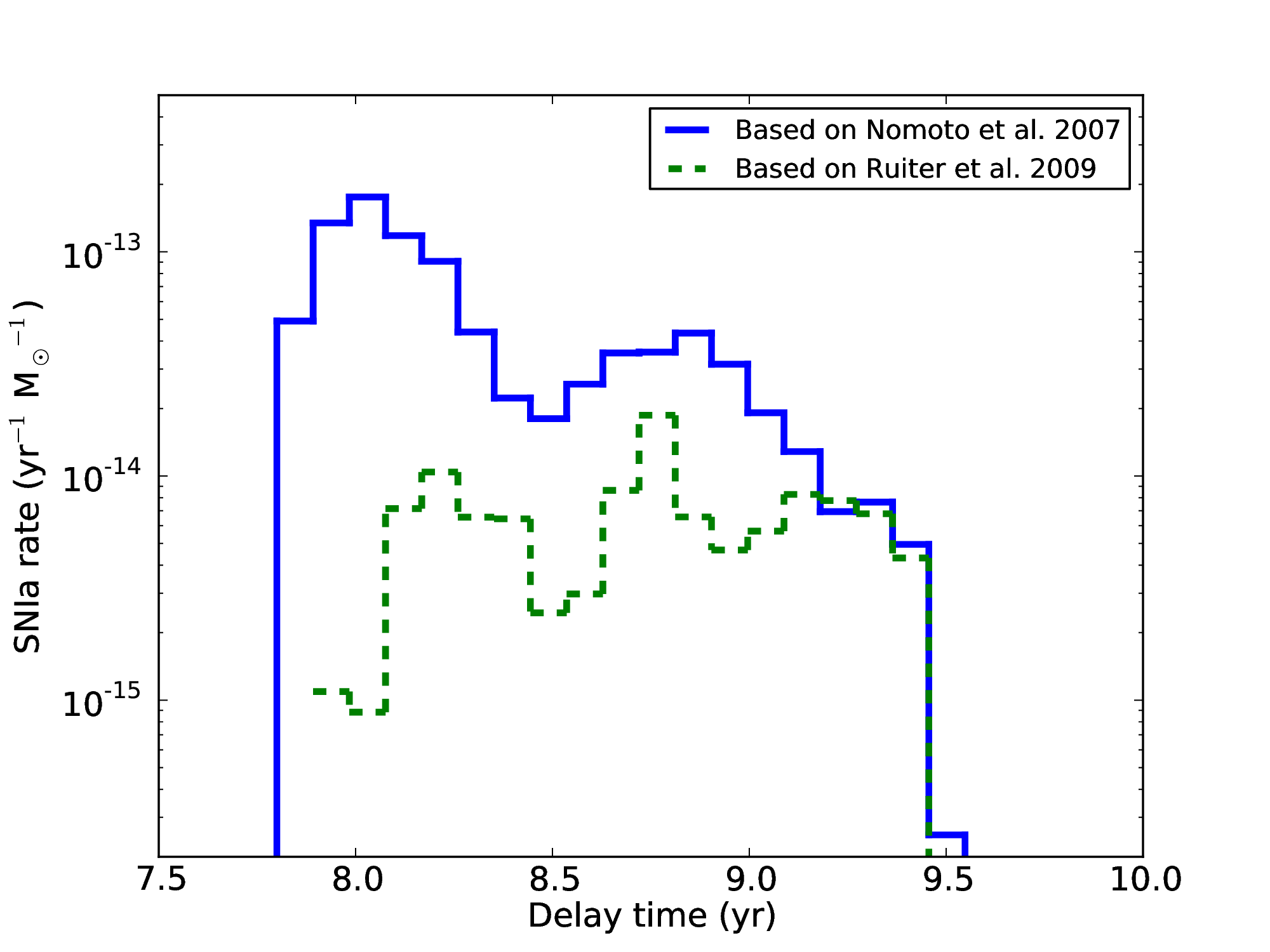}  \\
	a) DD channel & b) SD channel\\
\end{tabular}
\caption{\small SNIa rate per yr per M$_{\odot}$ formed stellar mass as a function of delay time. a) On the left rates from the DD channel. The delay time is in Gyr. The solid line represents a simulation using model $\mga$ for the common envelope evolution, dashed $\maa$. Overplotted with black circles are the observed values of the SNIa rate \citep[see][for a review]{Mao11b}. b) On the right rates from the SD channel. The delay time is in yr (logarithmic). Note that no SNeIa where found when the retention efficiency of \citet{Yun10} are used.}
\label{fig:dtd}
\end{figure}

The DTD of model $\gamma\alpha$ and model $\alpha\alpha$ are similar in showing strong declines with time, approximately $\propto 1/t$, see Fig. \ref{fig:dtd}a. This is expected when the delay time is dominated by the merger time of a double WDs as a result of gravitational wave emission. Fig. \ref{fig:dtd}a shows that these mergers are expected to take place in young as well as old populations. In absolute numbers the DTD of model $\mga$ is similar to model $\maa$ as well; the time-integrated rates are $2.0\cdot 10^{-4}\ M_{\odot}^{-1}$ for model $\mga$ resp. $3.3 \cdot 10^{-4}\ M_{\odot}^{-1}$ for model $\maa$. \looseness-1

The recovered DTD from a compilation of observations and method by \citet{Mao11b} has a similar shape as our synthetic DTDs, but has a normalization that is a factor $\sim$10 higher. The integrated rate from \citet{Mao10, Mao11b} is $18-29 \cdot 10^{-4}\ M_{\odot}^{-1}$ resp. $>34 \cdot 10^{-4}\ M_{\odot}^{-1}$ based on galaxy cluster measurements and cluster iron abundances respectively. However, recent measurements in field galaxies have shown lower rates; $0.6-1.0 \cdot 10^{-4}\ M_{\odot}^{-1}$  by \citet{Per12} as rescaled by \citet{Mao12}, $13\pm 1.5 \cdot 10^{-4}\ M_{\odot}^{-1}$ by \citet{Mao12} and $5 \pm 2\cdot 10^{-4}\ M_{\odot}^{-1}$ by \citet{Gra12} (however \citet{Gra12} state this value is likely an lower limit). At this moment it is unclear if the different observed integrated rates are due to systematic effects or if there is a real enhancement of SNeIa in cluster galaxies. See also \citet{Mao12} for a discussion. 

The normalisation of the synthetic SNIa rates is dependent on assumptions in the simulation; for example the stellar winds, CE evolution, assumed binary fraction and the initial distribution of masses and orbital parameters. However, preliminary results show that the integrated rates are not affected by factors as large as $\sim$ 10 (see also Claeys et al. in prep.). Therefore the synthetic rates are inconsistent with the rate of \citet{Mao11} and the main contribution to the SNIa rate has to come from other channels. On the other hand, the synthetic rates are consistent with the lower limit of the SNIa rate in field galaxies. Future observations will help to solve this enigma.

\section{Single degenerate channel}
The single degenerate scenario explains in a very natural way the uniformity in the SNIa lightcurves, as all SD events occur when the WD approaches the Chandrasekhar mass. The theoretical SD rates predicted by different BPS codes, however, vary over four orders of magnitude and do not agree with each other or with observational data \citep{Nel12}. The exact origin of these differences remains unclear so far.  

\begin{figure*}
    \centering
    \begin{tabular}{c c}
	\includegraphics[angle=90,width=0.5\columnwidth]{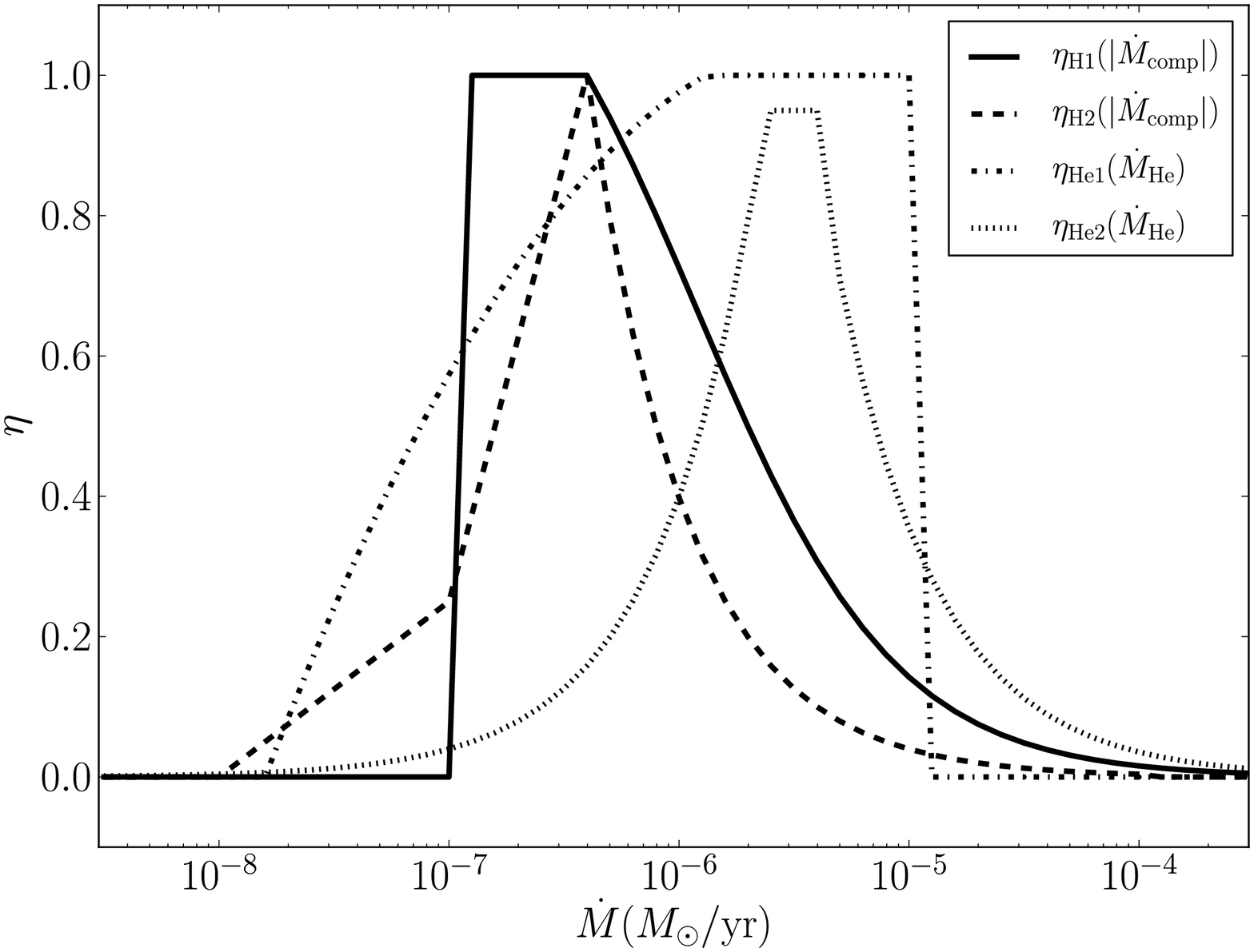} &
	\includegraphics[angle=90,width=0.5\columnwidth]{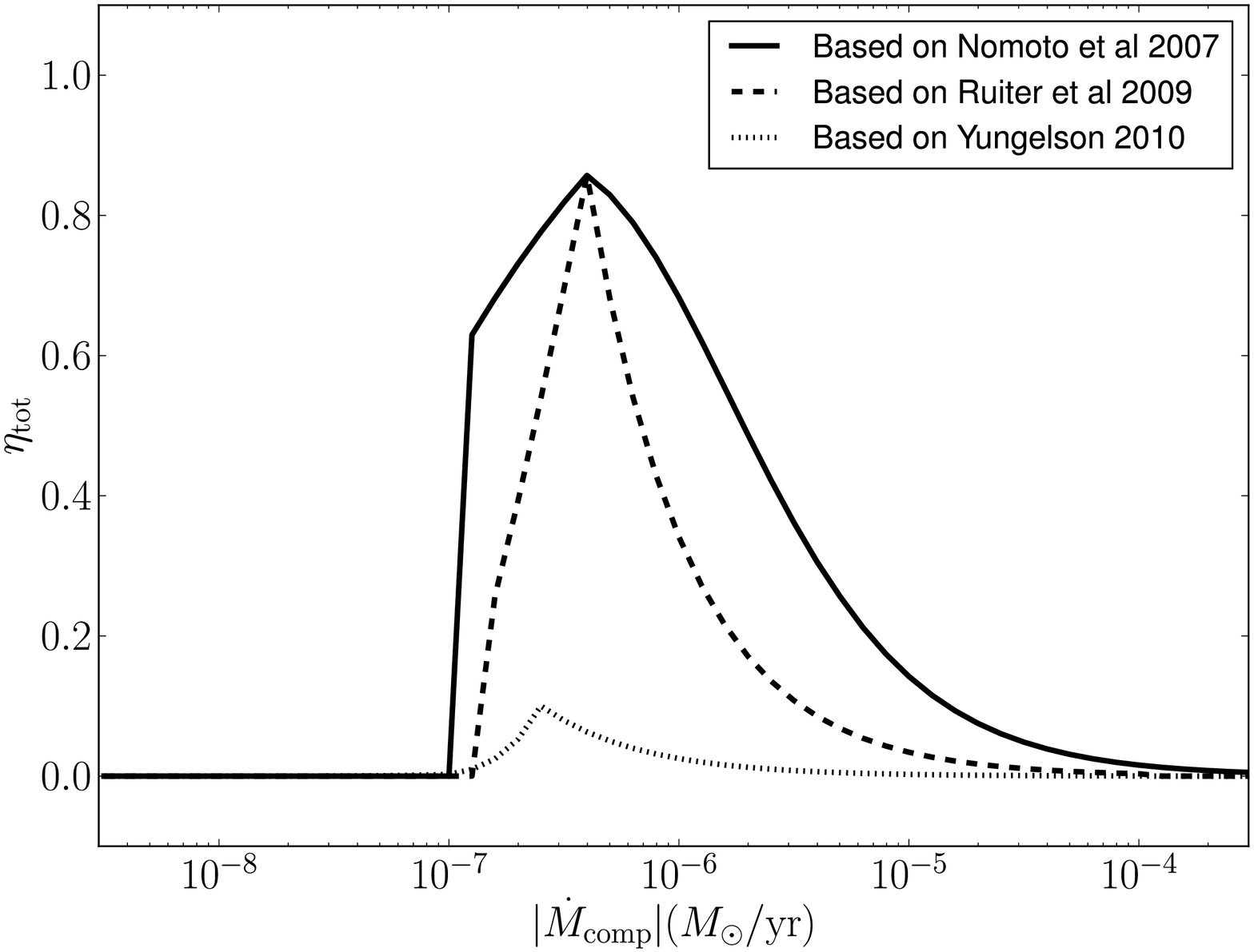} \\
	a) & b) \\
	\end{tabular}
    \caption{\small The mass retention efficiency of WD hydrogen accretion as a function of the mass transfer rate. $\eta_{\rm tot}$ represents the fraction of the transferred mass that is retained by the WD.  In the figure we have assumed the white dwarf to be of one solar mass. a) Left the hydrogen retention efficiency $\eta_\mathrm{H1}$ (\citet{Nom07, Hac08} - solid line) \& $\eta_\mathrm{H2}$ (\citet{Rui09}, \citet{Yun10} - dashed line) and the helium retention efficiency $\eta_\mathrm{He1}$ (\citet{Hac99, Kat99} - dot-dashed line) \& $\eta_\mathrm{He2}$ (\citet{Ibe96} - dotted line) are shown. b) Right the total retention efficiency based on \citet{Nom07, Rui09, Yun10} are shown resulting from different combinations of the hydrogen and helium retention efficiencies.}  
    \label{fig:SD}
    \end{figure*}

A crucial phase in the evolution of SD systems is the phase of mass transfer and accretion onto the WD. The efficiency of mass accretion, hereafter retention efficiency $\eta$, is poorly understood because of processes such as novae and stable burning\footnote{Note that the retention efficiency is of little importance to the DD channel and that the predicted DTDs from different BPS codes are comparable \citep{Nel12}.}. If $\eta=0$ no mass is retained by the WD, when $\eta=1$ the WD accretes matter conservatively. In BPS codes different prescriptions \citep{Nom07, Rui09, Yun10} are used for the retention efficiency. The retention efficiency of hydrogen accretion that is burned into helium is shown in Fig. \ref{fig:SD}a, solid $\eta_\mathrm{H1}$ for \citet{Nom07} and dashed $\eta_\mathrm{H2}$ as assumed by \citet{Rui09} and \citet{Yun10}. At mass transfer rates $\dot M<10^{-7} \mathrm{M}_{\odot} \mathrm{yr}^{-1}$ novae occur taking away a large fraction of the accreted mass. At mass transfer rates of a few times $10^{-7} \mathrm{M}_{\odot} \mathrm{yr}^{-1}$ hydrogen burns stably into helium on the surface of the WD. At higher mass transfer rates, strong winds develop from the surface of the WD. In the wind regime $\eta_\mathrm{H1} > \eta_\mathrm{H2}$ because a stripping effect of the donor by the wind of the WD \citep{Hac08} is taken into account in $\eta_\mathrm{H1}$. After a fraction of the hydrogen-rich matter is burned into helium-rich material, a fraction of the helium-rich material is burned into carbon-rich matter. For the efficiency of helium retention $\eta_\mathrm{He}$ two prescriptions exist as well. The dot-dashed line in Fig. \ref{fig:SD}a shows $\eta_\mathrm{He}=\eta_\mathrm{He1}$ as assumed by \citet{Nom07} and \citet{Rui09} and $\eta_\mathrm{He2}$ assumed by \citet{Yun10}.

The total retention efficiencies $\eta_\mathrm{tot} = \eta_\mathrm{H}(\dot{M}_\mathrm{H}) \cdot \eta_\mathrm{He}(\dot{M}_\mathrm{He})$ after hydrogen and helium burning is shown in Fig. \ref{fig:SD}b. This figure shows a large variety in the assumed retention efficiencies, where the most optimistic case is given by \citet{Nom07} and pessimistic by \citet{Yun10}. We have used the BPS code SeBa to simulate the SNIa rates in the SD channel for each retention efficiency. The simulated SNIa rates are significantly affected by which prescription for the mass retention efficiency is used. In Fig. \ref{fig:dtd}b the resulting DTDs are shown assuming a $\gamma$-CE. The integrated rates vary between $5.8\cdot 10^{-5}\ M_{\odot}^{-1}$,  $1.9 \cdot 10^{-5}\ M_{\odot}^{-1}$ and an upper limit of $1 \cdot 10^{-7}\ M_{\odot}^{-1}$ when the retention efficiency is used as in \citet{Nom07}, \citet{Rui09} and \citet{Yun10} respectively. 
This study shows that depending on the assumptions for the retention efficiency the contribution to the SNIa rate from the SD channel can be negligible \citep[for][]{Yun10} to moderate (for \citealt{Nom07} and \citealt{Rui09}, where the difference in the integrated rates is of the order of a few). Predictions from BPS codes on the contribution from the SD channel to the SNIa rate are dependent on many assumptions e.g. CE evolution, but is crucially dependent on the retention efficiency. For more information about this study, see Bours, Toonen, Nelemans in prep.

\section{A comparison of binary population synthesis codes}
In the previous sections we have discussed the effect on the SNIa rate of two phases in binary evolution that are not well understood; the CE and mass accretion onto WDs. However, we are not able to reproduce the same DTDs and progenitors as other BPS codes when assuming their assumptions for the CE and $\eta$. To understand the differences in the predictions of the various BPS codes, we started a collaboration to compare four BPS codes. The goal is to investigate whether differences in the simulated populations are due to numerical effects, or whether they can be explained by differences in the input physics. Regarding the latter, we aim to identify and qualify which assumptions in BPS codes affect the results most, and hence should be studied in more detail.

The codes involved are the Binary\_c code \citep{Izz04, Izz06, Izz09, Cla11}, the Brussels code \citep{DeD04, Men10}, SeBa \citep{Por96, Nel01, Too12} and Startrack \citep{Bel02,Bel08, Rui09}. The comparison focuses on the evolution of low- and intermediate-mass binaries containing one or more WDs. For this project the four codes have equalized their assumptions as much as possible, even when these assumptions are not assumed to be realistic. These assumptions are: \\
$\cdot$ The same initial distribution of primary mass, mass ratio and separations; \\
$\cdot$ The same CE prescription and efficiency;\\
$\cdot$ Stable mass transfer is conservative and to all types of stars;\\
$\cdot$ No tidal effects, magnetic braking or wind accretion.\\
Preliminary conclusions are that when the input assumptions are equalized, the different BPS codes give similar populations. There are small differences, however they are not caused by numerical differences, but can be explained by inherent differences in the input physics, most importantly: \\
$\cdot$ The initial-final mass relation for white dwarfs;\\
$\cdot$ For which systems does mass transfer occur stably, and when is it unstable?\\
$\cdot$ When mass transfer is stable, what is the mass transfer rate?\\
$\cdot$ How much mass (and angular momentum) is lost in the wind of stars? \\
$\cdot$ The evolution of hydrogen-poor helium-burning stars is very important for SNIa progenitors. Not much is known yet about these objects regarding e.g. the stellar evolution, winds or mass transfer stability.  \\

\acknowledgements This work was supported by the Netherlands Research Council NWO (grants \# 643.200.503, \# 639.042.813, \# 639.073.803, \# 614.061.608) and by the Netherlands Research School for Astronomy (NOVA).

\bibliographystyle{asp2010}
\bibliography{toonen}

\end{document}